\documentclass[aps,prb,twocolumn,showpacs]{revtex4}

\usepackage{graphics} \usepackage{graphicx} \usepackage{amssymb}
\usepackage{amsfonts} \usepackage{amsmath} \usepackage{bm}

\newcommand{\bea}{\begin{eqnarray}}
\newcommand{\eea}{\end{eqnarray}}
\newcommand{\beq}{\begin{equation}}
\newcommand{\eeq}{\end{equation}}
\newcommand{\benu}{\begin{enumerate}}
\newcommand{\enu}{\end{enumerate}}

\newcommand{\al}{\alpha}
\newcommand{\be}{\beta}
\newcommand{\ga}{\gamma}
\newcommand{\om}{\omega}
\newcommand{\Om}{\Omega}
\newcommand{\ep}{\epsilon}

\newcommand{\si}{\sigma}
\newcommand{\dl}{\delta}
\newcommand{\Dl}{\Delta}

\newcommand{\ham}{\mathcal{H}}

\newcommand{\ptl}{\partial}

\newcommand{\cda}{c^{\dagger}}
\newcommand{\bk}{{\bf k}}
\newcommand{\bq}{{\bf q}}
\newcommand{\bp}{{\bf p}}
\newcommand{\bQ}{{\bf Q}}
\newcommand{\br}{{\bf r}}

\begin{document}
\title{Quasiparticle mirages in the tunneling spectra of d-wave superconductors}
\date{\today}
\author{I. Paul$^{1}$, A. D. Klironomos$^{1,2}$ and M. R. Norman$^1$}
\affiliation{
$^1$Materials Science Division, Argonne National Laboratory, Argonne, IL 60439\\
$^2$Department of Physics, The Ohio State University, Columbus, Ohio 43210
}

\begin{abstract}
We illustrate the importance of many-body effects in the Fourier transformed local density of
states (FT-LDOS) of d-wave superconductors from a model of electrons
coupled to an Einstein mode with energy $\Om_0$.
For bias energies significantly larger than $\Om_0$
the quasiparticles have short lifetimes due to this coupling, and the FT-LDOS is featureless if the
electron-impurity scattering is treated within the Born approximation. In this regime
it is important to include boson exchange for the electron-impurity
scattering which provides a `step down' in energy for the electrons and allows for
long lifetimes. This many-body effect produces qualitatively different results,
namely the presence of peaks in the FT-LDOS which are mirrors of the quasiparticle
interference peaks which occur at bias energies smaller than $\sim \Om_0$.
The experimental observation of these quasiparticle mirages would be an important step forward
in elucidating the role of many-body effects in FT-LDOS measurements.
\end{abstract}

\maketitle

Many-body effects are known to influence the electron spectral function in cuprates,
in particular the peak-dip-hump lineshape seen in the superconducting state by both
angle resolved photoemission and tunneling~\cite{eschrigreview}.
The nature of the bosonic modes that give rise to this lineshape
is a topic of much debate.
Among the possibilities that have been discussed in the literature
are certain optical
phonons, as well as the spin resonance seen by inelastic neutron
scattering.  They all have a
comparable excitation energy, $\Om_0 \approx 40$ meV for optimal doped
Bi$_2$Sr$_2$CaCu$_2$O$_{8 + \dl}$ (Bi2212), which is also
near the energy of the antinodal gap, $\Delta_A$.
Very recently it has been suggested that scanning tunneling spectroscopy
(STS) can be useful in revealing the electron-boson coupling in the
cuprates~\cite{jlee}. In STS
the differential conductance $dI/dV(\br, eV)$ is a measure
of the local density of states (LDOS) $\rho (\br, \om = eV)$ of the electrons, while
the Fourier transform ($\br \rightarrow \bq $) yields
$\rho(\bq, \om)$ (FT-LDOS). The location of the peaks of $\rho(\bq, \om)$
provides information about the excitation spectrum of the
electronic states~\cite{hoffman,kyle}.
In view of the suggestion of Ref.~\onlinecite{jlee}, it is particularly important to carefully examine
how electron-boson coupling affects these peaks~\cite{balatsky}.

Modulations of the LDOS and the concomitant peaks in $\rho(\bq, \om)$
are due to electrons scattering from impurities.
In most studies
the electron-impurity scattering has been treated either within a
$T$-matrix approximation~\cite{wanglee}  or a Born approximation~\cite{capriotti}.
In this framework, the peaks in $\rho(\bq, \om)$ appear at wavevectors $\bq = \bk_f - \bk_i$, where
$\bk_f$ and $\bk_i$ are points of high density of states at the
energy $\om$. This approach has been quite successful in understanding FT-LDOS data for
$|\om| \lesssim \Dl_{A}$, and has been denoted as
the `octet' model~\cite{kyle}.
We note that the success of this approach depends on the existence
of electronic states with long lifetimes in this energy range.
Subsequent to this, there have been attempts to go beyond these approximations by including
the interaction of the electrons with either phonons or spin fluctuations~\cite{balatsky,polkovnikov}.

The purpose of this paper is to reveal that for bias energies
greater than $\sim \Delta_A$,
where no quasiparticles are observed in photoemission~\cite{arpes},
there can be sharp peaks in $\rho(\bq, \om)$
due to a boson exchange process that appears beyond the Born approximation for the electron-impurity
scattering.
For $|\om| > \Om_0$, an electron can decay into a lower energy state by
emitting a boson of energy $\Om_0$. As a result, the lifetime of the electrons with energy $\om$
significantly larger than $\Om_0$
is severely reduced, which is the primary reason for the absence of sharp peaks in $\rho(\bq, \om)$
at the level of the Born approximation.
But, going beyond the Born approximation in the process shown in Fig.~1b,
an electron first emits a boson which reduces its energy to $|\om| - \Om_0$.  If this reduced energy
is not significantly larger than $\Om_0$,  the resulting electron state
(the internal fermion line of the diagram) is once again long-lived,
and consequently contributes to peaks in
$\rho(\bq, \om)$ by scattering from the impurities.  In this process, the location of the peaks
at $\om$ is determined by the excitation spectrum at an energy
$(|\om| - \Om_0) \rm{sgn}(\om)$.
That is, one has mirages of the octet peaks at $|\om_{QI}| \lesssim \Dl_{A}$
that are mirrored at an energy $|\om_{QM}| = |\om_{QI}| + \Om_0$.
We note that this argument is based entirely on the energetics of the
electron-boson interaction, and does not depend on the momentum space structure of the bosons (i.e., whether
the bosons are spin fluctuations peaked at large wavevectors or phonons peaked at small
wavevectors), though of course the momentum form factor of the bosons will lead to quantitative
differences.

\begin{figure}
\centerline{\includegraphics[width=3.4in]{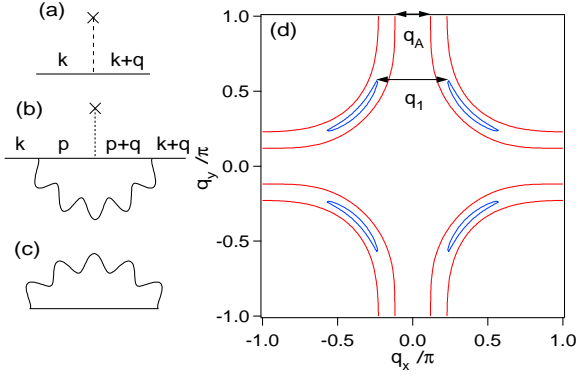}}
  \caption{(Color online) Panels (a)-(c) show the relevant diagrams considered in this paper: Born diagram, vertex correction, and the self-energy, respectively.
The cross represents an impurity and the wiggly line a boson.
  In (a) and (b) the electrons (straight lines) are dressed by the self-energy.
  In (c) the electron line is bare.
 Panel (d) shows the constant energy contours of the bare dispersion
 for two energies: $-\om = 20$ meV (the closed `banana' contour)
  and $-\om = 76$ meV (the open contours), noting that
$\Delta_A = 37$ meV. The wavevector
  ${\bf q}_1$ connects the tips of the closed contour,
and ${\bf q}_A$ connects the inner of the two open contours along the antinodal direction.}
\label{fig1}
\end{figure}

\emph{Model.} We study a two-dimensional system of superconducting electrons described
by a BCS model
interacting with a boson mode, and coupled to an isotropic potential scatterer.
It is described by the Hamiltonian
$\ham = \ham_{BCS} + \ham_{im} + \ham_{el-b}$. Here $\ham_{BCS} = \sum_{\bk, \si} \ep_{\bk} \cda_{\bk \si}
c_{\bk \si} + \sum_{\bk} \Dl_{\bk}( \cda_{\bk \uparrow} \cda_{- \bk \downarrow} +
c_{- \bk \downarrow} c_{\bk \uparrow})$, where $\cda_{\bk \si}$ ($c_{\bk \si}$) creates (annihilates)
electrons with spin $\si$ at wavevector $\bk$, the normal state dispersion is given by the tight
binding expansion
$\ep_{\bk} = t_0 + t_1 (\cos k_x + \cos k_y)/2 + t_2 \cos k_x \cos k_y + t_3 (\cos 2k_x + \cos 2k_y)/2
+ t_4 (\cos 2k_x \cos k_y + \cos k_x \cos 2k_y)/2 + t_5 \cos 2k_x \cos 2k_y$, and the superconducting gap has the d-wave form
$\Dl_{\bk} = \Dl_M (\cos k_x - \cos k_y)/2$, with the lattice constant set to unity
($M$ denoting the $(\pi,0)$ point).
In order to study the sensitivity
of $\rho(\bq, \om)$ to the dispersion,
we considered two sets of values for the parameters $t_i$,
one taken from Ref.~\onlinecite{eschrig} and the other from Ref.~\onlinecite{kaminski}.
For both dispersions, the antinode is at ${\bf k}_{A} = (1,0.18)\pi$.  They differ in that the first
dispersion has $\ep_M$ close
to the Fermi energy (-34 meV), whereas it is further away for the second (-119 meV).
We focus here on results from the second dispersion, though qualitatively similar
results were obtained from the first.
For $\Dl_{M}$, we choose 40 meV, a typical value for optimal doped Bi2212.

The electron-impurity scattering is given by
$\ham_{im} = V_0 \sum_{\bk, \bq, \si} \cda_{\bk + \bq \si} c_{\bk \si}$, where $V_0=1$eV in our calculation (note that $V_0$ simply sets the scale for the FT-LDOS).
For the sake of concreteness,
we take the coupling between the bosons and the electrons to be of the form
 $\ham_{el-b} = g \sum_i {\bf S}_i \cdot {\bf s}_i$, where ${\bf S}_i$ and ${\bf s}_i$ are
the spin fluctuation and the electron spin operators respectively at site $i$,
though our results hold equally well for optical phonons.
We fix the magnitude of the
coupling constant $g$ from the condition that in the normal state ($\Dl_{\bk} = 0$) the
inverse quasi-particle weight $z^{-1} = 1 - \frac{\ptl{\rm Re}\Sigma(\om)}{\ptl \om} \approx 2$ at the Fermi energy,
where $\Sigma (\om)$ is the electron self-energy due to interaction with the spin fluctuations.
This gives
$3g^2 = 0.0176$ eV$^2$.
The dynamics of the spin fluctuations is given by
$\chi_{\mu \nu}(\bQ, i \Om_n) = \chi(i \Om_n) \dl_{\mu \nu}$, with
$\chi( i \Om_n) =
 2 \Om_0/(\Om_n^2 + \Om_0^2)$, where
$\chi_{\mu \nu}(\bQ, \tau) = \langle T_{\tau} S_{\mu}(\bQ, \tau) S_{\nu}(- \bQ, 0) \rangle$ is the
spin fluctuation propagator (the overall prefactor being absorbed into the definition of $g$).
Here $\mu$, $\nu$ are spatial indices,
$\Om_n$ is a bosonic Matsubara frequency, and the mode energy $\Om_0$ is taken
to be 39 meV.
In our model, we take
the bosons to be independent of momentum for the
following reasons. First, it allows us to concentrate on the energetics.  As discussed
in Ref.~\onlinecite{eschrig}, as long as the form factor of the bosons is finite throughout the zone,
then the self-energy will be dominated by the density of states singularities associated with
the antinodal and $M$ points of the internal fermion line.
Second, it greatly simplifies the calculation of the vertex diagram (Fig.~1b).

\begin{figure}
 \centerline{\includegraphics[width=3.4in]{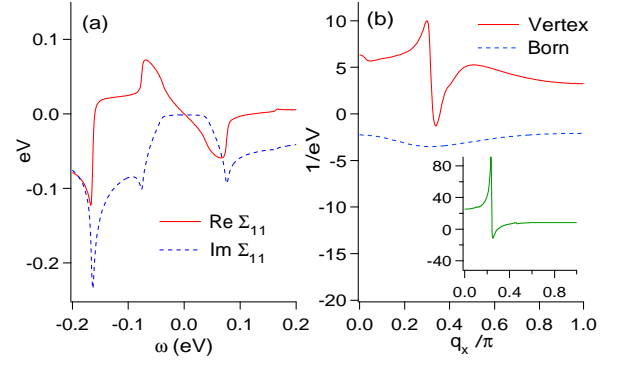}}
  \caption{(Color online) (a) The real and imaginary parts of the 11 component of the self-energy.
(b) Contribution to the FT-LDOS from the vertex diagram for $- \om = \Om_0+\Delta_{A} = 76$ meV.
 The dashed line shows the contribution from the Born diagram.  The inset shows the
  contribution from the Born diagram for electrons without the self-energy correction.}
\label{fig2}
\end{figure}

Using Nambu notation, the $2 \times 2$ electron Green's function with self-energy
correction (Fig.~1c) is given by
\[
G^{-1}(\bk, z )
=
\left[
\begin{array}{cc}
z - \ep_{\bk} - \Sigma_{11}(z)
&
- \Dl_{\bk} - \Sigma_{12}(\bk, z)
\\
- \Dl_{\bk} - \Sigma_{12}^*(\bk, z)
&
z + \ep_{\bk} - \Sigma_{22}(z)
\end{array}
\right],
\]
where $z$ is the complex frequency. The diagonal self-energies at T=0 are given by \cite{eschrig}
$\Sigma_{11/22}(z) = 3 g^2 \sum_{\bk } [z \pm \ep_{\bk}(1+ \Om_0/E_{\bk})]/
[z^2 - (\Om_0 + E_{\bk})^2] $, where $E_{\bk} = \sqrt{\ep_{\bk}^2 + \Dl_{\bk}^2}$.
In the above, the $\bk$-summation is performed numerically with an intrinsic
lifetime broadening factor $\eta = 2$ meV for $z= \om + i \eta$.
In Fig.~2a we show the real and imaginary components of $\Sigma_{11}(z)$ as a
function of $\om$.
${\rm Im} \Sigma_{11}(\om)$ is zero for $|\om| \lesssim \Om_0$, and has
pronounced peaks at $|\om| = \Om_0 + \Dl_{A}$ and $\Om_0 + E_M$.
Consequently, for energies up to and slightly beyond $\Om_0$, the electronic states form
well defined quasiparticles, while for $|\om| \gg \Om_0$ they do not.
Strictly speaking, for momentum independent bosons, the off-diagonal self-energies
vanish due to the $d$-wave symmetry. However, in order to keep the antinode
gap energy unrenormalized, we use the ansatz
$\Sigma_{12}(\bk, \om) = \Dl_{\bk} [Z({\om}) - 1]$, where
$Z(\om) = 1 - \frac{1}{2 \om}{\rm Re}
[ \Sigma_{11}(\om) + \Sigma_{22} (\om)]$. In this scheme, the renormalized
dispersion is
$\tilde{E}_{\bk} \approx [\Dl_{\bk}^2 + \ep_{\bk}^2/Z^2(\tilde{E}_{\bk})]^{1/2}$.

\emph{Results.} We first compute the contribution to the FT-LDOS from the Born
diagram (Fig.~1a). This is given by $
\rho^{(b)}(\bq, \om) = - (2/\pi) {\rm Im} [ V_0 \sum_{\bk}
G_{1 \al}^{R}(\bk, \om) (\hat{\tau}_3)_{\al \be} G_{\be 1}^{R}(\bk + \bq, \om)]$,
where summation over repeated Nambu indices $\al, \be = (1,2)$ is implied,
and $\hat{\tau}_i$ are Pauli matrices in Nambu space,
with $R$ denoting retarded propagators.
The variation of  $\rho^{(b)}(\bq, \om)$ versus $\bq=(q_x,0)$ is shown as the dashed curve in
Fig.~2b for $-\om = \Delta_A + \Om_0$.
We note that, due to large lifetime broadening,
there is no peak in $\rho^{(b)}(\bq, \om)$
at this energy. The effect of the lifetime can be seen clearly by comparing this curve
with the one in the inset which is obtained
by computing $\rho^{(b)}(\bq, \om = - \Om_0 - \Dl_{A})$
with $\hat{\Sigma} = 0$ (i.e., for unrenormalized electrons).

\begin{figure}
 \centerline{\includegraphics[width=3.4in]{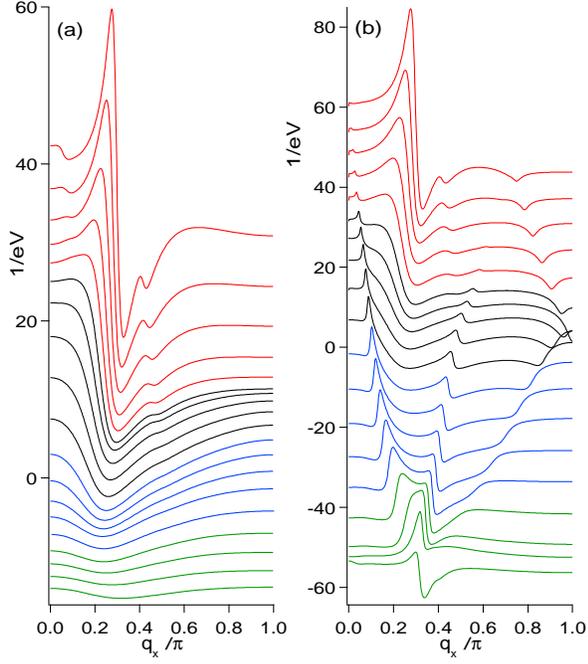}}
  \caption{(Color online) (a) Born contribution, and (b) Born plus vertex contributions,
to the FT-LDOS for energies ranging from near $-\Om_0$ (-40 meV, top curve) to
$-\Om_0 - \Delta_A$ (-76 meV, bottom curve) in steps of 2 meV.
The curves are offset for clarity.
}
  \label{fig3}
\end{figure}

Next we calculate the FT-LDOS contribution of the vertex diagram (Fig.~1b).  This is given by
\bea
\rho^{(v)}(\bq, \om) &=&
- \left(\frac{2}{\pi}\right) {\rm Im} \left[ V_0 \sum_{\bk}
G_{1 \al}^{R}(\bk, \om) T_{\al \be}^{R}(\bq, \om)
\right. \nonumber \\
&\times&
\left.
G_{\be 1}^{R}(\bk + \bq, \om) \right],
\eea
where
\bea
T_{\al \be}(\bq,  i \om_n)
&=&
 \frac{3 g^2}{\be} \sum_{\Om_n, \bp} \chi( i \Om_n)
G_{\al \ga}(\bp, i \om_n - i \Om_n)
\nonumber \\
&\times&
(\hat{\tau}_3)_{\ga \dl} G_{\dl \be}(\bp + \bq, i \om_n - i \Om_n).
\nonumber
\eea
Here $\be$ is the inverse temperature and $\om_n$ is a fermionic Matsubara frequency.
We note that in our model, the matrix $\hat{T}$ does not depend on $\bk$
due to the momentum independence of the bosons.
The variation of $\rho^{(v)}(\bq, \om)$ along the bond direction is plotted in
Fig.~2b for  $\om = - (\Om_0 + \Dl_{A})$.
We note the pronounced
structure with a sharp maximum at $q_x = 0.30 \pi$ and a sharp minimum at $0.34 \pi$.
As  $|\om|$ increases from $\om = - (\Om_0 + \Dl_{A})$ to   $- (\Om_0 + \tilde{E}_{M})$,
where $\tilde{E}_M$ is the renormalized energy at the $M$ point ($\sim$ 60 meV),
 this structure evolves into a single broad minimum at
$q_x = 0.23 \pi$ (not shown).  This can be contrasted with the Born result, where there is only
weak structure in this entire energy range.

\begin{figure}[!t]
 \centerline{\includegraphics[width=3.4in]{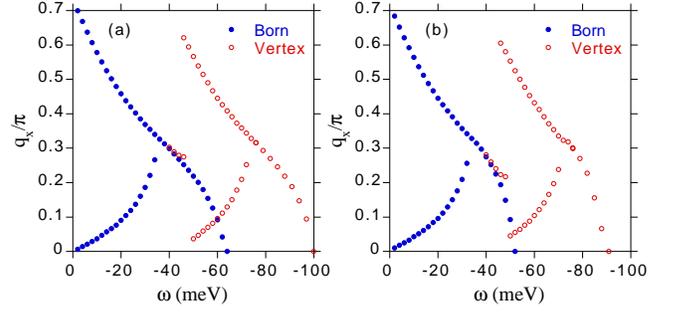}}
  \caption{(Color online) Dispersion of the the Fourier peaks along the  bond direction
versus energy for (a) the real part and (b) the imaginary part of the Born and vertex diagrams.
Note the characteristic `$\lambda$' shape of the Born dispersion and its $\Om_0$ (39 meV) displaced
`mirage' in the vertex dispersion.
}
\label{fig4}
\end{figure}

 For $|\om| \gg  \Om_0$, due to a large ${\rm Im} \Sigma_{\al \al}(\om)$,
 the $\bk$-summation in Eq.~(1) involving
the product of two $G$ functions yields a quantity (denoted as $C$) which is approximately constant
as a function of $\bq$.
The variation of $\rho^{(v)}(\bq, \om)$ with $\bq$ for these energies is mainly due to
that of $T_{\al \be}^{R}(\bq, \om)$. For the frequency summation  involved in the
computation of $T_{\al \be}^{R}(\bq, \om)$, the main contribution  is due to the
bosonic pole which puts the electrons with momentum $\bp$ and $\bp + \bq$
(Fig.~1b) at an
energy $\om + \Om_0$ for $\om$ negative.
This contribution can be written as
(for $\om < -\Om_0$)
$
\hat{T}^{R}(\bq, \om)_{{\rm coh}}
= 3 g^2 \sum_{\bp} \hat{G}(\bp, \om + \Om_0) \hat{\tau}_3 \hat{G}(\bp + \bq,
\om + \Om_0).
$
This expression is reminiscent of the Born contribution at an energy $\om+\Om_0$,
and gives rise to sharp structure for $|\om| - \Om_0 \lesssim \Om_0$ since
the electronic states are well defined at these energies.

To understand this in greater detail, we plot in Fig.~3a the contribution to the FT-LDOS
 from the Born diagram,
and in Fig.~3b the sum of the Born and vertex contributions,
versus $(q_x,0)$, for $\Om_0 < |\om| < \Om_0 + \Delta_A$.  In Fig.~4, we in turn plot the resulting
peak dispersions from the real and imaginary parts of the Born and
vertex diagrams for all $|\om| < \Om_0 + \tilde{E}_M$.
We note that the Born dispersion is well defined in
the energy range between 0 and $\tilde{E}_M$, although the Born peaks are damped
for $\Om_0 < |\om| < \tilde{E}_M$ due to lifetime broadening.  For $|\om| < \Delta_A$, there are two peaks.
The one at larger $q_x$  corresponds to scattering between the tips of the of the constant
energy contours, which look like bananas in this energy range.  This is denoted by
the vector ${\bf q}_1$ in Fig.~1d (structure corresponding to the even larger
vector ${\bf q}_5$ of Ref.~\onlinecite{kyle} is not plotted in Fig.~4).
The one at smaller $q_x$ corresponds to the so-called Tomasch peak noted in
Ref.~\onlinecite{capriotti}.
It corresponds to where the banana first stops overlapping its $q_x$ displaced image.
For $\Delta_A < |\om| < \tilde{E}_M $,
one finds a dominant maximum which traces out the separation of the inner contours in Fig.~1d along the antinodal ($(\pi,0)-(\pi,\pi)$) direction (denoted as ${\bf q}_A$ in Fig.~1d),
with secondary peaks (not plotted in Fig.~4) corresponding to connecting an inner to an outer
contour or an outer to an outer contour (these secondary peaks have less weight due to the
reduced spectral weight of the outer contours).  The combination of these peaks
(two for $|\om| < \Delta_A$ and one for $|\om| > \Delta_A$)
gives a characteristic `$\lambda$' shape to the overall bond oriented dispersion, as is obvious
from Fig.~4.
Note there are some differences in the dispersions associated with the real and
imaginary parts of the Born diagram and their connection to the vectors denoted in
Fig.~1d.  This is due to several factors:
  the finite lifetime of the electronic states, the dispersion $\epsilon_k$,
  and the fact that Re$G$ has a zero where Im$G$ has a pole.

We now turn to the vertex diagram.  Its dispersion (Fig.~4) essentially mirrors the Born dispersion
at a bias energy displaced by $\Om_0$.  As a consequence, we denote this as a `quasiparticle
mirage'.  In addition, for biases near $\Om_0$, the energy undisplaced Born term is also reflected
 in the vertex diagram (since the external lines in Fig.~1b have well defined spectral peaks,
and the boson exchange process has a limited phase space, for these energies).
We note that there are some differences in the lineshapes of the imaginary
 part of the vertex
 term as compared to that of the Born term displaced by $\Om_0$, as is evident in Fig.~3.
 This occurs since both the real and imaginary
  parts of the components of $\hat{T}$ contribute to $\rho^{(v)}(\bq, \om)$ as
  $C$ is a complex quantity.

  Next, we comment on a few qualitative aspects of the vertex contribution to the FT-LDOS.
  (i) The inclusion of a momentum form factor for the boson propagator should not
  make any qualitative change to the peak structure. Such a form factor
  (peaked around some $\bQ_0$) can be thought of as a momentum
  constraint forcing $\bk \approx \bp + \hat{\alpha} \bQ_0$ (where $\hat{\alpha}$ is a lattice
  group operation). However, in the mechanism discussed
  above, the electrons with momentum $\bk$ and $\bk + \bq$ (external lines of Fig.~1b)
  do not play any special role.
  (ii) It is important that the boson that provides the
  `step down' in energy has a sharp spectral function. A finite lifetime of the boson, or its dispersion
 with $q$,
  will broaden the Fourier peaks.
  For similar reasons, we anticipate that higher order vertex corrections will lead to
  weaker and broader contributions to the Fourier peaks because of the additional momentum
  sums involved.
  (iii) Recently, the observation of  peaks in the
  Fourier transformed $d^2I/dV^2$ spectrum at
  $\bq \approx (0.4 \pi, 0)$ has been reported for
  optimal doped Bi2212 at $|\om| \approx \Om_0 + \Dl_{A}$~\cite{jlee}.  This is comparable
  to the peak position we find from our vertex corrected calculation at this bias energy.
 So far, though, no dispersion of these peaks has been reported.

  \emph{Conclusion.} We demonstrated the importance of vertex corrections to electron
  impurity scattering in the study of tunneling spectroscopy data for the cuprate
  superconductors for absolute bias energies larger than $\Om_0$,
where $\Om_0$ is the excitation
  energy of a boson coupled to the electrons.
  The vertex correction is due to emission and re-absorption of a boson
  by the electrons which leads to a `step down' of the internal fermion line to energies
  where quasiparticle states are well defined, which as a consequence gives
rise to sharp peaks in the FT-LDOS.  We denote these new peaks as `quasiparticle mirages',
 whose dispersion mirrors the previously observed quasiparticle interference peaks
 at absolute biases smaller than $\Om_0$.  The observation of these dispersive `mirages'
 would be an important reflection of the nature of the many-body interactions in cuprates.

This work was supported by the U.~S.~Dept.~of Energy, Office of Science,
under Contract No.~DE-AC02-06CH11357.
Part of the calculations were performed at the Ohio Supercomputer Center
thanks to a grant of computing time.
The authors would like to thank
Mohit Randeria for suggesting this work.

\end{document}